\documentclass[twocolumn,apl,epsf,superscriptaddress]{revtex4}

\usepackage{amsmath}
\usepackage{amsfonts}
\usepackage{epsfig}
\usepackage{epsf}
\usepackage{array}
\usepackage{color}
\usepackage{ulem}

\begin{document}
\title{Transition-Metal Oxide (111) bilayers}
\altaffiliation{
Copyright  notice: This  manuscript  has  been  authored  by  UT-Battelle, LLC under Contract No. DE-AC05-00OR22725 with the U.S.  Department  of  Energy.   
The  United  States  Government  retains  and  the  publisher,  by  accepting  the  article  for  publication, 
acknowledges  that  the  United  States  Government  retains  a  non-exclusive, paid-up, irrevocable, world-wide license to publish or reproduce the published form of this manuscript, 
or allow others to do so, for United States Government purposes.  
The Department of Energy will provide public access to these results of federally sponsored  research  in  accordance  with  the  DOE  Public  Access  Plan 
(http://energy.gov/downloads/doe-public-access-plan)}

\author{Satoshi Okamoto}
\altaffiliation{okapon@ornl.gov}
\affiliation{Materials Science and Technology Division, Oak Ridge National Laboratory, Oak Ridge, Tennessee 37831, USA}
\author{Di Xiao}
\affiliation{Department of Physics, Carnegie Mellon University, Pittsburgh, Pennsylvania 15213, USA}

\begin{abstract}
Correlated electron systems on a honeycomb lattice have emerged as a fertile playground to explore exotic electronic phenomena. 
Theoretical and experimental work has appeared to realize novel behavior, including quantum Hall effects and valleytronics, 
mainly focusing on van der Waals compounds, such as graphene, chalcogenides, and halides.   
In this article, we review our theoretical study on perovskite transition-metal oxides (TMOs) as an alternative system to realize such exotic phenomena. 
We demonstrate that novel quantum Hall effects and related phenomena associated with the honeycomb structure could be artificially designed by such TMOs 
by growing their heterostructures along the [111] crystallographic axis. 
One of the important predictions is that such TMO heterostructures could support two-dimensional topological insulating states. 
The strong correlation effects inherent to TM $d$ electrons further enrich the behavior. 
\end{abstract}


\maketitle

\section{Introduction}

Two dimensional electron systems on a honeycomb lattice have been providing fascinating phenomena. 
A unique aspect of the honeycomb lattice is the appearance of Dirac points in the band structure, 
and its topological property is controlled by the gap opening at the Dirac points. 
Haldane first realized that introducing complex electron hopping amplitudes on a honeycomb lattice could realize the quantum Hall effect (QHE) 
in the absence of Landau levels \cite{Haldane1988}. 
This originates from the nontrivial band topology induced by the time reversal symmetry (TRS) breaking, which opens the gap at the Dirac point. 
Later, Kane and Mele realized that such nontrivial band topology could be induced without breaking the TRS 
when the relativistic spin orbit coupling (SOC) exists \cite{Kane2005}, 
opening the new field of topological insulators (TIs). 

Graphene, a monolayer of graphite, offers an ideal system to support such novel phenomena \cite{Novoselov2005,Zhang2005}. 
While the SOC in graphene turned out to be too small to realize TI states at accessible temperatures \cite{Min2006}, 
early theoretical work has predicted a novel integer QHE \cite{Zheng2002}, which has been experimentally confirmed \cite{Novoselov2005,Zhang2005}. 
More recently, a fractional QHE was also realized experimentally \cite{Bolotin2010}. 
Further interesting phenomena may be realized if the correlation effects are strong. 
It has been argued that graphene has a potential to realize novel ``chiral'' superconductivity by carrier doping \cite{Baskaran2002,BlackSchaffer2007,Pathak2010,Nandkishore2012}. 
In addition to graphene, other van der Waals compounds have also been attracting attention \cite{Bhimanapati2015}. 
These van der Waals compounds could contain transition-metal elements and, therefore, 
might be suitable for exploring phenomena associated with correlation effects, such as magnetism. 

In light of correlation effects, transition-metal oxides (TMOs) have a long history, covering magnetism, high-critical temperature ($T_c$) superconductivity, 
colossal magnetoresistance effects, and correlation-induced metal-insulator transitions.\cite{Imada1998}
Iridium-based oxides are one of the central focuses of recent studies 
because the energy scales of the Coulomb repulsive interaction and the SOC are compatible \cite{Kim2008,Pesin2010}. 
In their seminal work, Shitade and coworkers have performed density functional theory (DFT) calculations and 
proposed that Na$_2$IrO$_3$, having a honeycomb lattice formed by Ir atoms, would become a two-dimensional TI \cite{Shitade2009}. 
On the other hand, Chaloupka and coworkers took an alternative approach from a Mott insulating side \cite{Chaloupka2010} and
proposed that the low-energy behavior of Na$_2$IrO$_3$ is governed by the so-called Kitaev-Heisenberg model,\cite{Kitaev2006}
which is a candidate for realizing $Z_2$ quantum spin liquid (SL) states. 
However, later experimental measurements confirmed a magnetic long-range order in Na$_2$IrO$_3$ \cite{Singh2010,Liu2011,Ye2012}. 
To account for the experimental results, refined theoretical models were developed.\cite{Rau2014,Yamaji2014,Sizyuk2014} 

While Na$_2$IrO$_3$ turned out to be a trivial insulator with complex magnetic ordering, 
TMOs in general have great potential to serve as a major playground to explore novel phenomena originating from the strong SOC and correlation effects. 
Here, we would like to focus on artificial heterostructures of TMOs rather than bulk compounds. 
This is motivated by the recent development in thin-film growth techniques of TMOs. 
A variety of TMO heterostructures with atomic precision have been synthesized and analyzed \cite{Hwang2012}. 
TMO heterostructures have great tunability over fundamental physical parameters, 
including the local Coulomb repulsion, SOC, and carrier concentration. 
However, the effect of correlations to possible novel phenomena 
near Mott insulating states with a strong SOC remains to be explored. 

In this article, we review our theoretical work on the novel electronic property of TMO heterostructures. 
In contrast to the standard heterostructures grown along the [001] crystallographic axis, 
we consider those grown along an unconventional axis, i.e., [111] direction. 
This allows us to modify the underlying  lattice geometry from tetragonal to trigonal. 
By using multiple theoretical techniques, we demonstrate the correlated graphene-like behavior arising from such TMO heterostructures.


\section{Main idea}

\begin{figure}
\begin{center}
\includegraphics[width=0.95\columnwidth]{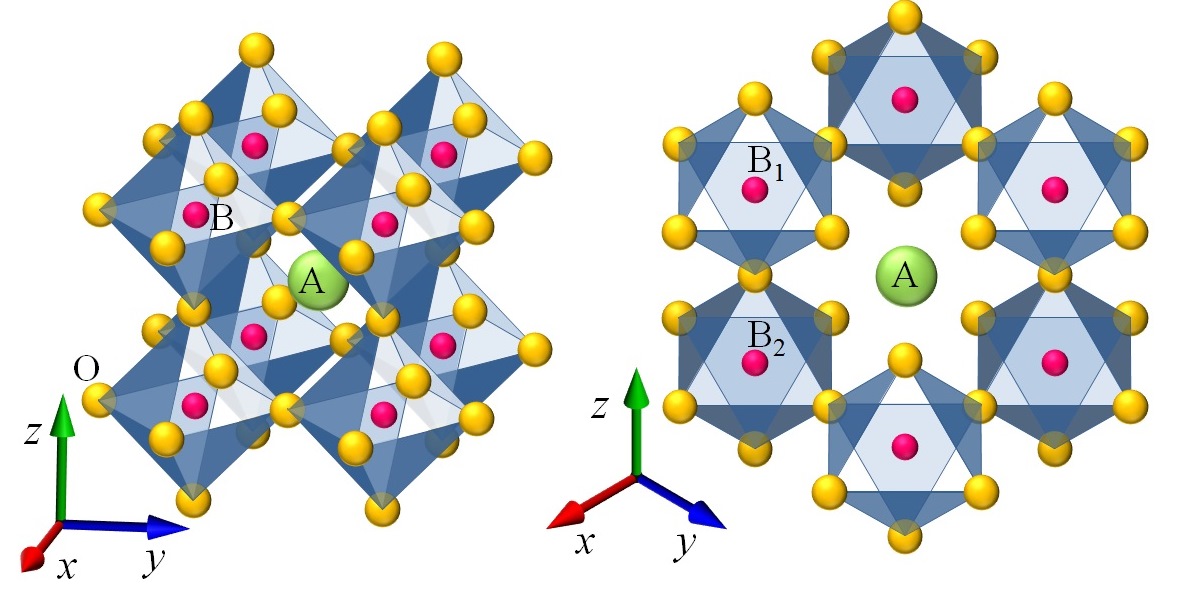}
\caption{(Color online) Crystal structure. 
Left: cubic perovskite ABO$_3$.
When a bilayer of ABO$_3$ is grown along the cubic [111] direction, 
it forms a buckled honeycomb lattice with two B sites, B$_1$ on the top layer and B$_2$ on the bottom layer (right). 
B$_1$ and B$_2$ could become inequivalent when substrate and capping materials are different or when a gate voltage is applied along the [111] direction. 
}
\label{fig:structure}
\end{center}
\end{figure}

Throughout this work, we consider perovskite TMOs, which have the chemical formula $AB$O$_3$. 
$A$ is normally a divalent alkaline earth element or a trivalent rare earth element, $B$ is a TM element, and O is oxygen. 
As shown in Fig. \ref{fig:structure}, TM $B$ sites form a cubic lattice in the ideal bulk perovskite. 
A crucial observation is that TM $B$ sites form a triangular lattice in a (111) plane and, then, 
two neighboring (111) planes of $B$ sites form a ``buckled'' honeycomb lattice \cite{Xiao2011}. 
Thus, one could anticipate ``Dirac points'' in their dispersion relations, just like graphene. 
In reality, it would be extremely difficult to create freestanding (111) bilayers. 
As in conventional (001) heteostructures, (111) bilayers must be grown on a substrate and capped properly. 
To maintain the inversion symmetry, the same material must be used for the substrate and the capping layer. 
The inversion symmetry can be intentionally broken by using different materials for the substrate and the capping layer 
or by applying a gate voltage along the [111] direction. 

\begin{figure}
\begin{center}
\includegraphics[width=0.8\columnwidth]{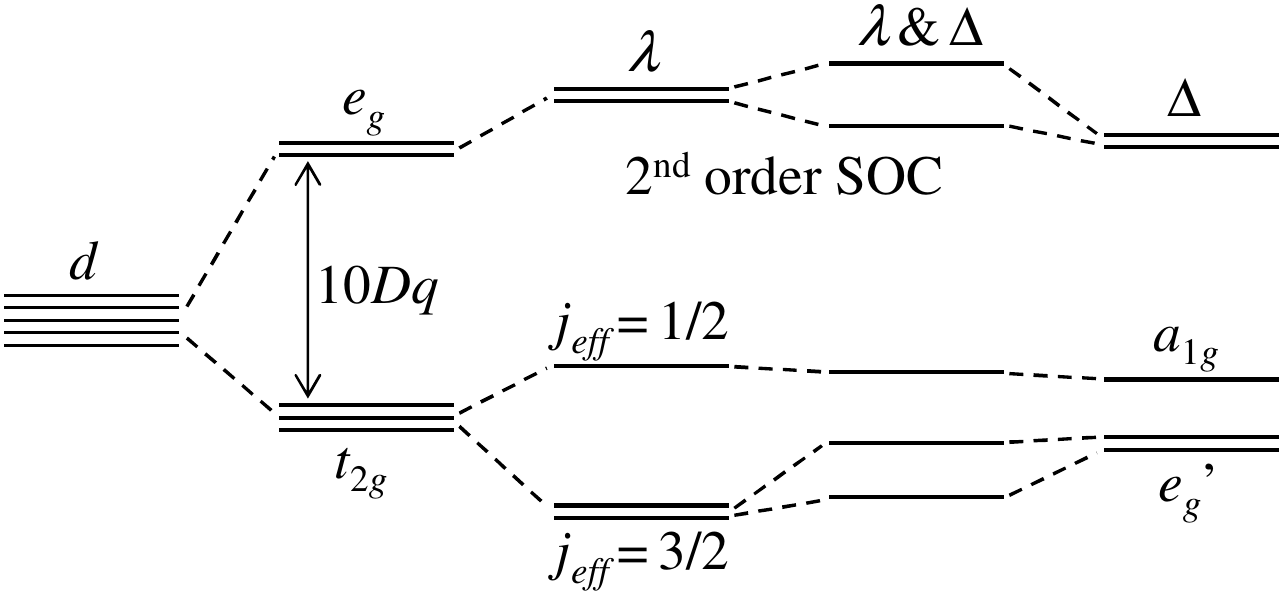}
\caption{Single-particle energy level for $d$ electrons.
$d$ level is degenerate in the spherical symmetric environment. 
Under the cubic environment, this splits into the triply degenerate $t_{2g}$ states and the doubly degenerate $e_g$ states with the level separation called $10Dq$.
With the SOC $\lambda$ is turned on,  the $t_{2g}$ level splits into the doubly degenerate $j_{eff}=1/2$ states and the four-fold degenerate $j_{eff}=3/2$ states, 
but the $e_g$ level  does not split while the $t_{2g}$-$e_g$ separation is increased by the second-order SOC. 
The trigonal crystalline field $\Delta$ alone does not affect the $e_g$ degeneracy either, 
while $t_{2g}$ level splits into the doubly degenerate $a_{1g}$ states and the doubly degenerate $e_g'$ states.  
When both $\Delta$ and $\lambda$ are turned on, all the degeneracy is lifted except for the Kramers degeneracy. }
\label{fig:level}
\end{center}
\end{figure}

\begin{figure}
\begin{center}
\includegraphics[width=0.95\columnwidth]{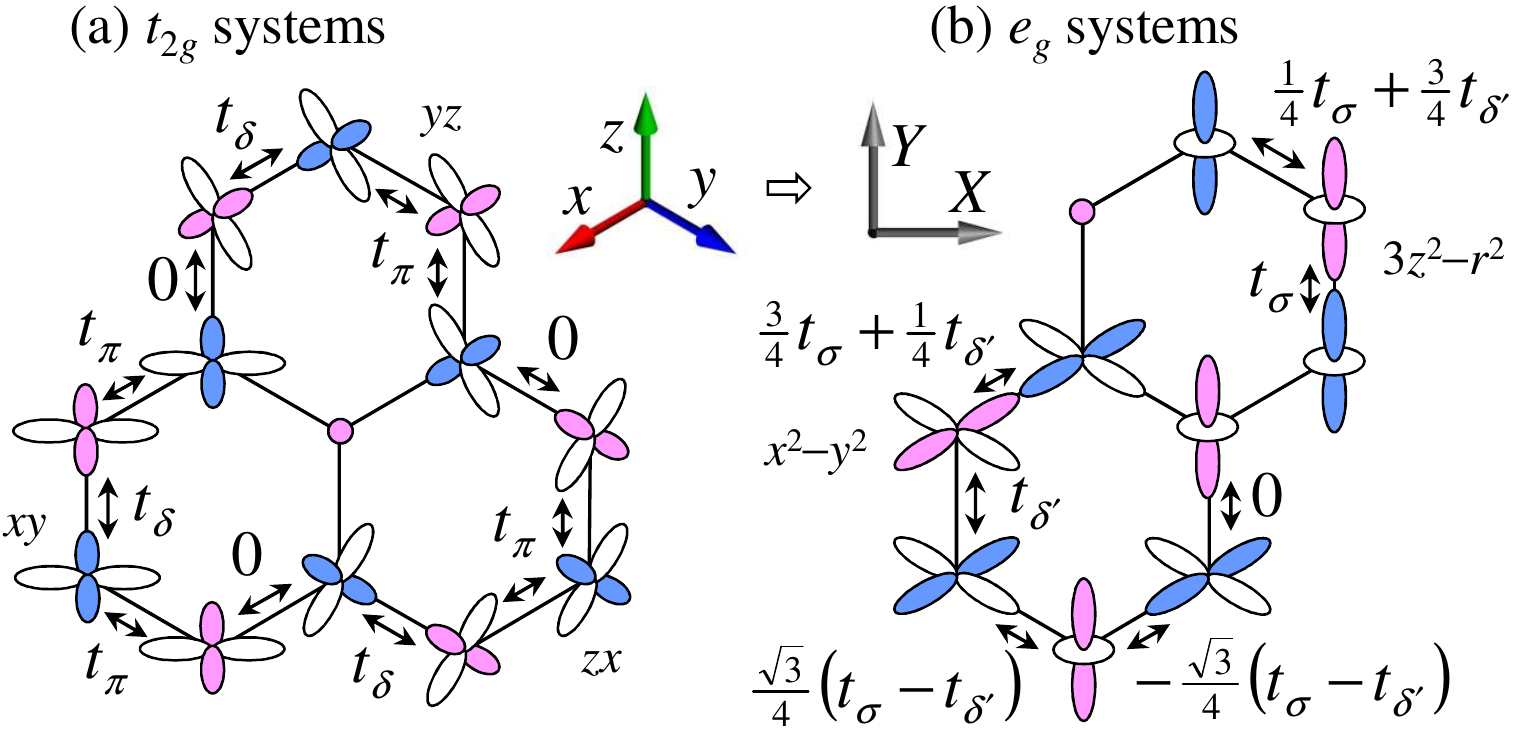}
\caption{(Color online) Nearest-neighbor transfer integrals for $t_{2g}$ systems (a) and $e_g$ systems (b). 
Different prefactors originate from the Slater-Koster parametrization 
considering the hybridization between TM $d$ orbitals and oxygen $p$ orbitals.
}
\label{fig:transfer}
\end{center}
\end{figure}
Can we align the Fermi level and Dirac points? 
Even if this is achieved, is the SOC strong enough to open a gap to induce a TI state? 
These are fundamental issues to realize TIs and could be overcome by properly choosing $A$ and $B$ elements. 
For the strong SOC, heavy $B$ elements are certainly favored, such as $4d$ or $5d$ TMs rather than $3d$ TMs. 
Integer filling required for TIs could be realized relatively easily. 
If necessary, doping concentration could be tuned by partially replacing $A$ site  ions by other ions with different valence. 
Also, gating could control doping concentration externally. 
By replacing $A$ site ions between the two (111) triangular lattices of $B$ ions by other ions with the same valence but different ionic radius, 
the strain value (distance between $B$ sites and crystal field) could be controlled. 
Strain could also be changed by replacing substrate and capping materials. 

One of the benefits of such (111) bilayers is the reduced symmetry of the crystalline field from octahedral ($O_h$) to trigonal ($C_{3v}$). 
The octahedral crystalline field splits the TM $d$ orbitals into three-fold degenerate $t_{2g}(yz, zx, xy)$ levels and 
two-fold degenerate $e_g(3z^2-r^2, x^2-y^2)$ levels well separated by the so-called $10Dq$ on the order of 3 eV.  
The trigonal crystalline field $\Delta$ further splits the $t_{2g}$ manifold into non-degenerate $a_{1g}$ and two-fold degenerate $e_g'$, and  
the SOC $\lambda$ splits the $t_{2g}$ manifold into two-fold degenerate $j_{eff}=1/2$ states and four-fold degenerate $j_{eff}=3/2$ states including Kramer's degeneracy. 
On the other hand, each effect alone does not break the degeneracy in the $e_g$ manifold 
while the SOC shifts the entire $e_g$ level upward by $\propto \lambda^2/10Dq$, i.e., second-order in the SOC strength. 
When both the trigonal field and the SOC are present, the $e_g$ degeneracy is lifted, 
resulting in the full degeneracy lifting except for the Kramer's degeneracy \cite{Vallin1970,Chen2009} as shown in Fig. \ref{fig:level}. 
Note that the level splitting within the $e_g$ manifold is proportional to $\tilde \lambda \propto \lambda^2 \Delta/(10Dq)^2$ and vanishes when $\Delta$ vanishes. 
Thus, TI states could be realized at multiple integer fillings in (111) TMO bilayers.

\section{Theoretical results}

This section consists of four subsections. 
In the first two subsections, we consider non-interacting or weakly interacting systems to explore possible materials realization of TIs. 
In the third subsection, we examine the correlation effects by means of  the dynamical mean field theory (DMFT).\cite{Georges1996} 
In the fourth subsection, we use a strong coupling approach from a Mott insulating regime. 
By these multiple approaches, we uncover novel properties of (111) TMO bilayers in a wide parameter space. 

\subsection{Tight-binding modeling}

\begin{figure*}[t]
  \begin{center}
    \includegraphics[width=1.5\columnwidth]{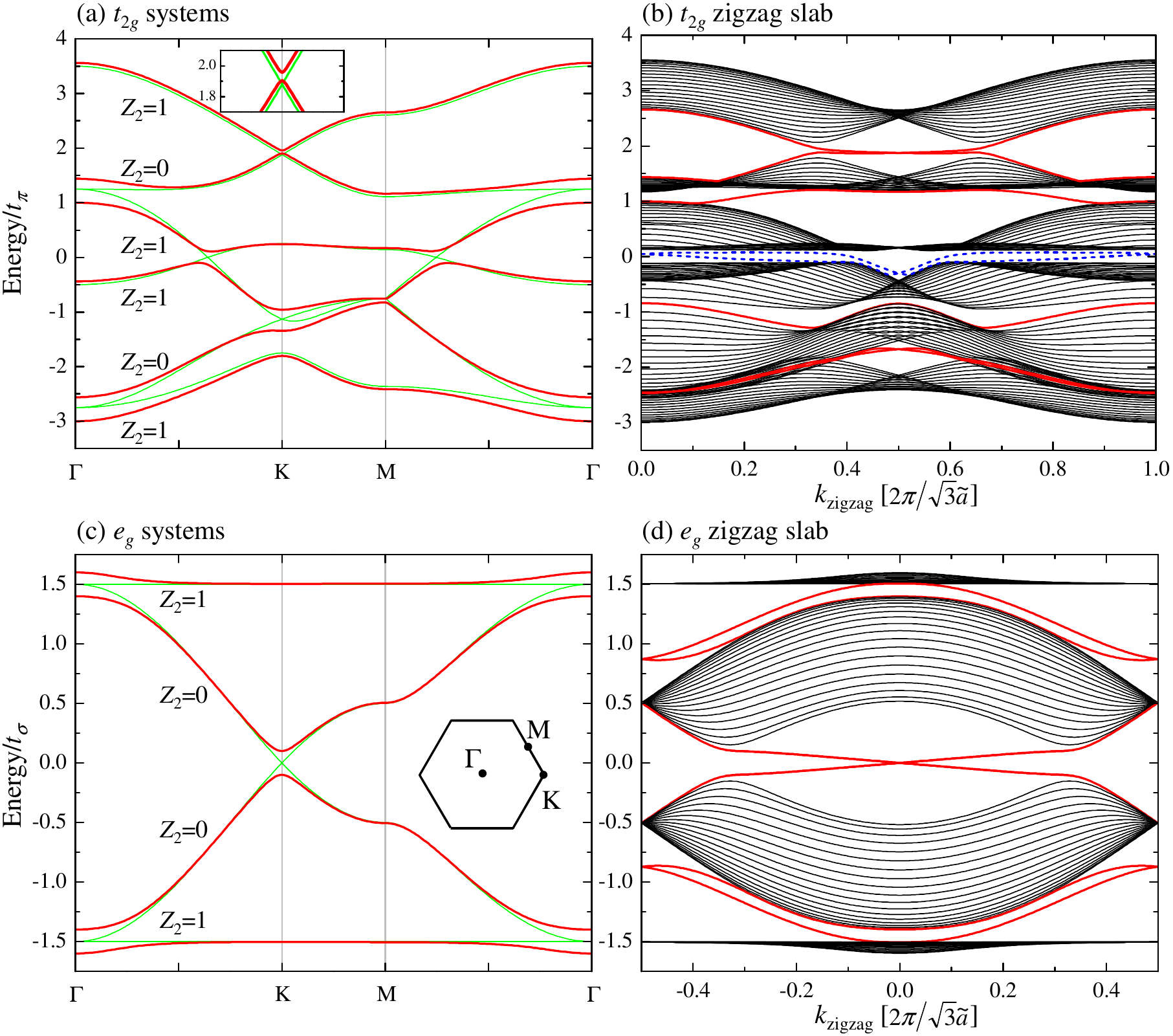}
    \caption{(Color online) 
Tight-binding results of the (111) bilayers. 
(a,b) $t_{2g}$ electron systems and (c,d) $e_g$ electron systems. 
(a) and (c) are dispersions for bulk (111) bilayers, while (b) and (d) are dispersions for finite-thickness zigzag slabs. 
Small hopping parameters $t_\delta$ and $t_{\delta'}$ are neglected for simplicity. 
Parameters are taken as $\Delta/t_\pi = 0.5$ with $\lambda/t_\pi = 1.5$ (thick lines) and $\Delta/t_\pi = 1.5$ with $\lambda/t _\pi= 0$ (thin lines) in (a), 
and $\tilde \lambda/ t_\sigma= 0.2$ (thick lines) and $\tilde \lambda / t_\sigma= 0$ (thin lines) in (b). 
Finite SOC gives rise to a nontrivial band topology as shown by band-dependent $Z_2$ indexes 
and gapless edge modes supporting the spin current (thick lines in the right figures).  
In (b), there are four edge channels between the third and the fourth bands as shown as dash lines.  
These edge channels do not support the spin current, consistent with $Z_2=0$ when the Fermi level is inside this gap. 
$\tilde a$ is the nearest-neighbor bond length projected on the (111) plane, i.e., $\tilde a = \sqrt{2/3}a$ with $a$ being the lattice constant of cubic perovskite. 
The figures ares adopted and modified from Ref. \cite{Xiao2011}. 
The inset of (c) is the first Brillouin zone of the hexagonal lattice. 
      } \label{fig:TB}
  \end{center}
\end{figure*}

With the structural consideration in the previous section, we first construct tight-binding models for $t_{2g}$ systems and $e_g$ systems. 
By noticing the relation between the cubic coordinate and hexagonal coordinate, 
it is straightforward to fix the transfer matrices, which depend on the direction and the pair of orbitals. 
Typical nearest-neighbor transfer integrals are shown in Fig. \ref{fig:transfer}. 
These are $d$-$d$ hopping through oxygen $p$ orbitals in between. 
Different prefactors originate from the Slater-Koster parametrization\cite{Slater1954} 
considering the hybridization between TM $d$ orbitals and oxygen $p$ orbitals.

Thus, for $t_{2g}$ electron systems, we may have 
\begin{equation}
H_{t_{2g}} =  -  \sum_{\stackrel{\scriptstyle \langle \vec r \vec r' \rangle \sigma}{\scriptstyle o o' \in t_{2g}}} 
\Bigl\{ t_{\vec r \vec r'}^{o o'} d_{\vec r o \sigma}^\dag d_{\vec r' o' \sigma}+ H.c. \Bigr\} \ + H_{t_{2g}}^{SOC}  + H_{t_{2g}}^{tri}
\end{equation}
and, for $e_g$ electron systems, 
\begin{equation}
H_{e_g} = - \sum_{\stackrel{\scriptstyle \langle \vec r \vec r' \rangle \sigma}{\scriptstyle o o' \in e_g}} 
\Bigl\{ t_{\vec r \vec r'}^{o o'} d_{\vec r o \sigma}^\dag d_{\vec r' o' \sigma}+ H.c. \Bigr\} 
+ H_{e_g}^{SOC}. 
\end{equation}
The SOC and the trigonal crystal field for $t_{2g}$ systems are given by 
\begin{equation}
H_{t_{2g}}^{SOC}= \lambda \sum_{\vec r} \vec l_{\vec r} \cdot \vec s_{\vec r}, 
\end{equation}
and
\begin{equation}
H_{t_{2g}}^{tri}= \Delta \sum_{\stackrel{\scriptstyle \vec r \sigma}{\scriptstyle o \ne o'}} d_{\vec r o \sigma}^\dag d_{\vec r o' \sigma}, 
\end{equation}
respectively. 
Here, $\vec l$ is the $l=2$ angular momentum operator projected on the $t_{2g}$ multiplet, and $\vec s$ is the spin operator. 
%
The second-order SOC for $e_g$ electron systems under the $C_{3v}$ symmetry is given in a compact form by 
\begin{equation}
H_{e_g}^{SOC} =  - \frac{\tilde \lambda}{2} \sum_{\stackrel{\scriptstyle \vec r \sigma}{\scriptstyle o o' \in e_g}}
d^\dag_{\vec r o \sigma} \tau^y_{o o'} \sigma^z_{\sigma \sigma} d_{\vec r o' \sigma}. 
\end{equation}
Here, $\tau^y$ is the Pauli matrix acting in the orbital space, and the uniform shift proportional to $\lambda^2/10Dq$ is absorbed in the chemical potential. 
The spin quantization axis is taken along the [111] crystallographic axis or perpendicular to the plane. 
Thus, in contrast to the $t_{2g}$ model, spin along this quantization axis is conserved and serves as a good quantum number as in the Kane-Mele model \cite{Kane2005}.

Using these TB models, we first examine their topological properties to determine the possible candidate materials for 2D TIs. 
Because of the multi-orbital nature of TMOs, our model gives rise to a very rich behavior of the topological band structure in the parameter space.  
For $t_{2g}$ systems, there is a competition between the SOC and the trigonal field, and 
depending on the relative strength between the two, the system falls into two different phases.  
We found that the highest two bands have robust topological properties irrespective to these parameters. 
Figures \ref{fig:TB} (a) shows the dispersion relations with 
$\Delta/t = 0.5$ and $\lambda/t = 1.5$ (red) and with $\Delta/t = 1.5$ and $\lambda/t = 0$ (green).
We determined the $Z_2$ topological invariant 
by evaluating directly from the bulk band structure \cite{Fu2006,Fukui2007} and by counting the number of edge states [Fig. \ref{fig:TB} (b)].  
By inspection, we find that $t_{2g}^2$, $t_{2g}^4$ and $t_{2g}^5$  systems are possible candidates for TIs, while $t_{2g}^1$ would become a topological semimetal.  
For $e_g$ systems, 
we found that $e_g^1$, $e_g^2$ and $e_g^3$ systems are possible candidates for TIs [see Fig.~\ref{fig:TB} (c) and (d) ].

It is worth mentioning that the lowest or highest band in the $e_g$ model is fairly flat. 
This originates from the fact that $x^2-y^2$ orbitals have very small transfer matrix elements along the $z$ direction $t_{\delta'}$; 
in Fig. \ref{fig:TB} (c),  $t_{\delta'}$ is taken to be zero, and non-zero dispersion comes from the SOC. 
It turned out that, when the TRS is broken by a magnetic field or spontaneous uniform magnetic ordering, i.e. FM, 
these bands have nonzero Chern number. 
Thus, in addition to TIs, the $e_g$ model could realize quantum anomalous Hall effects at $e_g^1$ or $e_g^3$ filling. 
Furthermore, when these bands are partially filled and long-ranged Coulomb interactions are present, 
fractional QHE could be realized as pointed out in Refs. \cite{Tang2011,Sun2011,Neupert2011,Venderbos2011,Wang2011b}.

\subsection{Density functional theory analyses}

Based on the TB results, we now look for candidate materials. 
While $t_{2g}^{2,4}$ and $e_g^{1,3}$ are also possible candidate systems, 
here we focus on $t_{2g}^5$ and $e_g^2$ systems because topological properties are robust.
For example, $t_{2g}^{2,4}$ systems would tend to become topological semimetals due to the band overlap, 
and $e_g^{1,3}$ systems could become unstable against the Jahn-Teller effect or $3x^2-r^2/3y^2-r^2$-type orbital ordering. 
For oxide thin-film growth, SrTiO$_3$ and LaAlO$_3$ are often used as insulating substrates. 
Based on this, we consider a TM $B$ ion to have the formal valence $+4$ in the form of Sr$B$O$_3$ or $+3$ in La$B$O$_3$. 
This limitation greatly reduces the number of materials combination. 

\begin{figure*}[t]
\begin{center}
\includegraphics[width=1.9\columnwidth]{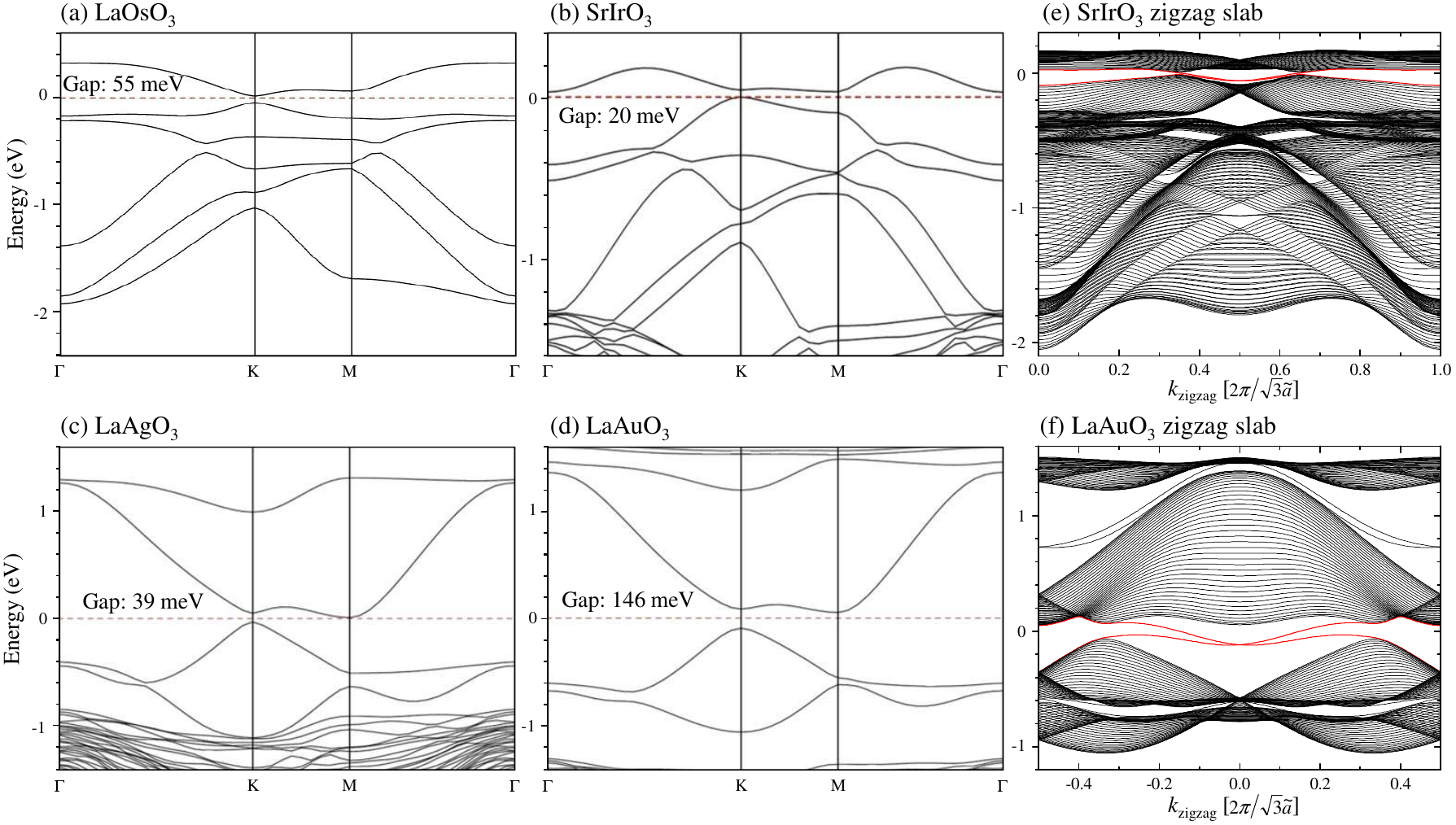}
\caption{(Color online) 
Density functional theory results of the dispersion relations of the (111) bilayer of TMOs. 
(a) LaOsO$_3$, (b) SrIrO$_3$, (c) LaAgO$_3$, and (d) LaAuO$_3$ are for bulk (111) bilayers, and 
(e) SrIrO$_3$ and (f) LaAuO$_3$ are for finite-thickness zigzag slabs. 
Bilayers shown in (a), (c), and (d) are grown between LaAlO$_3$, 
while that in (b) is grown between SrTiO$_3$. 
These systems are topological insulators with the gap amplitude indicated. 
The gappless edge modes at zero energy in (e) and (f) confirm the non-trivial band topology.  
The Fermi level is taken to be 0 of the vertical axis. 
Figures (a-d) are adopted and modified from Ref. \cite{Xiao2011}. 
Figures (e) and (f) are taken from Ref. \cite{Okamoto2014}.}
\label{fig:DFT}
\end{center}
\end{figure*}

For $t_{2g}^5$ systems, we come up with 
LaRu$^{3+}$O$_3$, LaOs$^{3+}$O$_3$, SrRh$^{4+}$O$_3$ and SrIr$^{4+}$O$_3$, and 
for $e_g^2$ electron systems
LaAg$^{3+}$O$_3$ and LaAu$^{3+}$O$_3$. 
While these materials have not been synthesized with the perovskite structure except for SrIrO$_3$ \cite{Cao2007}
(for example, according to Ref.~\cite{Ralle1993}, LaAuO$_3$ has CaF$_2$ structure rather than the perovskite), 
it is worth examining all these systems to prove our theoretical idea. 
We hope that advanced crystal synthesis techniques will allow fabrication of our predicted structure or 
that other systems with similar crystal and electronic structures will be discovered in the near future. 

We examined these (111) bilayers using DFT methods. 
DFT calculations were performed using the projector augmented wave method\cite{Kresse99}
with the generalized gradient approximation in the parametrization of Perdew, Burke and Enzerhof\cite{Perdew96} for exchange correlation 
as implemented in the Vienna Ab Initio Simulation Package\cite{VASP}. 
The detail of our DFT calculations can be found in Ref. \cite{Xiao2011}.

Figure \ref{fig:DFT} summarizes our DFT results. 
LaRuO$_3$ and SrRhO$_3$ (not shown) turned out to be topological semimetals, rather than TIs, because of the overlap between the conduction bands and the valence bands. 
Other candidate systems are found to become TIs with the Fermi level inside the nontrivial gap. 
To demonstrate their nontrivial band topology, 
we have derived Wannier functions \cite{Wannier} to construct finite-thickness zigzag slabs of SrIrO$_3$ (111) bilayer and LaAuO$_3$ (111) bilayer. 
As shown in Fig. \ref{fig:DFT} (e) and (f), these (111) bilayers have gapless edge modes crossing the Fermi level.

\subsection{Dynamical mean field theory analyses}

In the previous subsection, we did not consider correlation effects which could modify the band structure results. 
For example, correlation effects could induce some symmetry breaking such as ferromagnetic or antiferromagnetic (AFM) ordering. 
Furthermore, a Mott transition (correlation-induced metal to insulator transition) could take place. 
Here, we examine these two possibilities using DFT supplemented DMFT \cite{Georges1996}, i.e., DFT+DMFT.\cite{Okamoto2014} 
We focus on SrIrO$_3$ (111) bilayer and LaAuO$_3$ (111) bilayer because 
the former is already known to exist and the latter has the largest nontrivial gap.

The DMFT calculations were carried out using a single-particle Hamiltonian $H_{single}$ and a many-body part $H_U$. 
$H_{single}$ is parametrized in the Wannier basis.\cite{Wannier} 
We express $H_U$ in terms of real orbitals, either $t_{2g}$ or $e_g$, as\cite{Sugano70} 
\begin{eqnarray}
H_U \!\!\! &=& \!\!\! U\sum_o d^\dag_{o \uparrow} d_{o \uparrow} d^\dag_{o \downarrow} d_{o \downarrow}
+U' \sum_{o \ne o'} d^\dag_{o \uparrow} d_{o \uparrow} d^\dag_{o' \downarrow} d_{o' \downarrow} \nonumber \\
&&\! \!\! + (U'-J) \sum_{o > o'} \Bigl( d^\dag_{o \uparrow} d_{o \uparrow} d^\dag_{o' \uparrow} d_{o' \uparrow} 
+ d^\dag_{o \downarrow} d_{o \downarrow} d^\dag_{o' \downarrow} d_{o' \downarrow}  \Bigr) \nonumber \\
&&\!\!\! + J \sum_{o \ne o'} \Bigl( 
d^\dag_{o \uparrow} d_{o' \uparrow} d^\dag_{o' \downarrow}d_{o \downarrow}
+ d^\dag_{o \uparrow} d_{o' \uparrow} d^\dag_{o \downarrow} d_{o' \downarrow} \Bigr), 
\label{eq:Coulomb}
\end{eqnarray}
where $o$ and $o'$ stand for $(yz,zx,xy)$ for SrIrO$_3$ and $(3z^2-r^2,x^2-y^2)$ for LaAuO$_3$.  
$U$ and $U'$ are the intraorbital Coulomb interaction and the interorbital Coulomb interaction, respectively, 
and $J$ represents the interorbital exchange interaction (fourth term) and the interorbital pair hopping (fifth term). 
For orbitals with the $t_{2g}$ or $e_g$ symmetry, $U'=U-2J$. 

Impurity problems for DMFT are solved by using a finite temperature exact diagonalization technique \cite{Caffarel94,Perroni07,ARPACK}. 
One of the advantages of the exact diagonalization impurity solver is the direct access to the spectral function without employing a maximum-entropy analytic continuation. 
The detail of our DFT+DMFT calculations can be found in Ref. \cite{Okamoto2014}.

Are SrIrO$_3$ and LaAuO$_3$ in AF trivial insulating phases or TI phases? 
If they are in AF trivial phases, can we turn them into TI phases by suppressing magnetic ordering? 
To answer these questions, we estimate realistic Coulomb interactions by using the constrained random phase approximation
(cRPA)\cite{Aryasetiawan2004,Kozhevnikov2010}. 
For SrIrO$_3$ bilayer, we took the Slater parameters $F_{0,2,4}$ for Sr$_2$IrO$_4$ from Ref.~\cite{Arita2012} and 
deduced $U$ and $J$  as 2.232 eV and 0.202 eV, respectively, for the $\{xy, yz, zx\}$ basis. 
For LaAuO$_3$ bilayer, 
we directly computed these parameters for the $\{3z^2-r^2,x^2-y^2\}$ basis, 
%
%
%
%
and the resultant $U$ and $J$ are $U=1.80$ eV and $J=0.225$ eV, respectively.  

\begin{figure}
\begin{center}
\includegraphics[width=0.9\columnwidth]{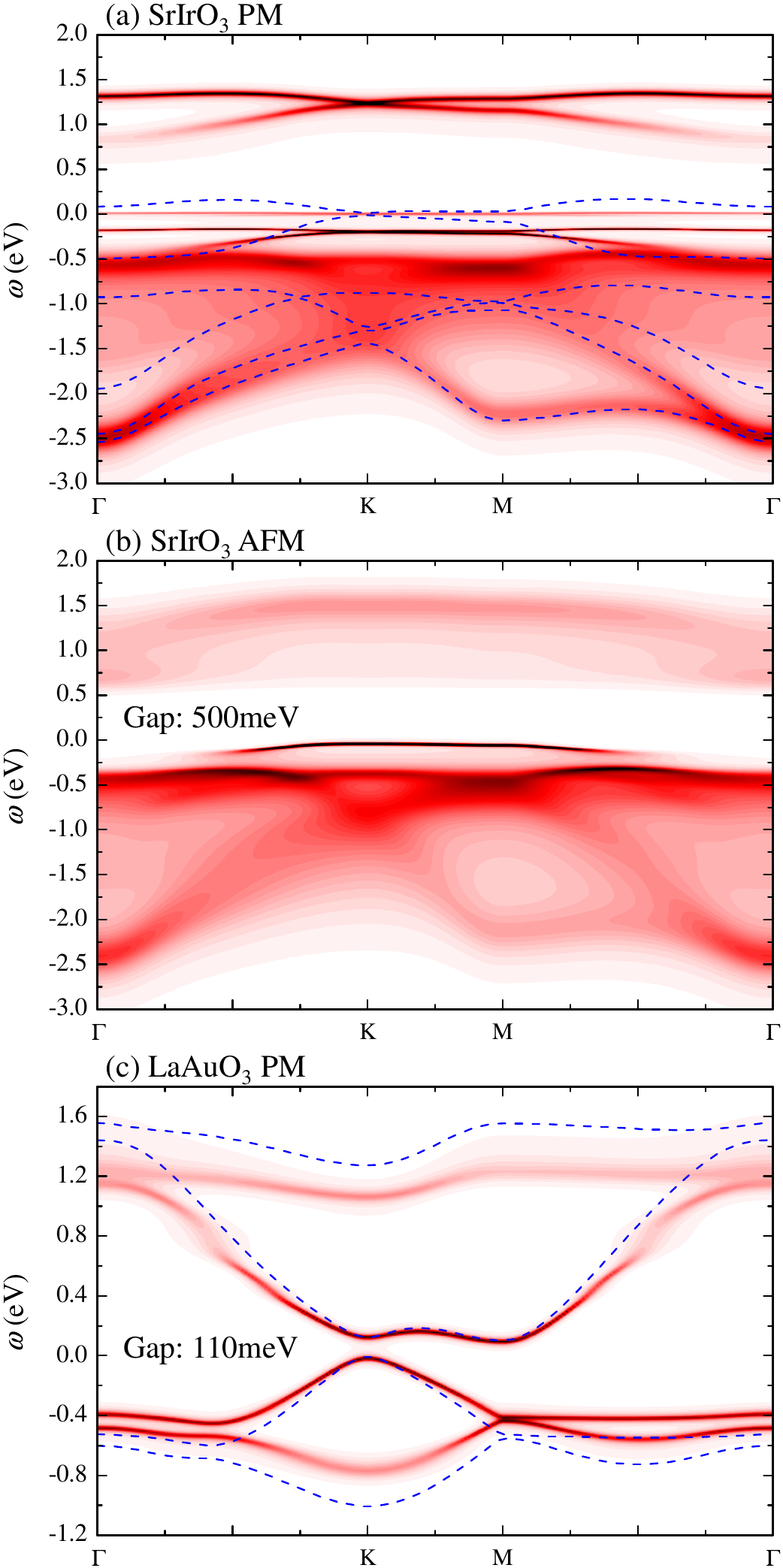}
\caption{Color online) 
DFT+DMFT result of the ARPES spectra of bulk SrIrO$_3$ (111) bilayer in (a) PM and (b) AFM phases and  bulk LaAuO$_3$ (111) bilayer (c). 
Thin dashed lines are DFT (Wannier) dispersion curves. 
Four lowest dispersions dominated by $j_{eff}=3/2$ states are downshifted by $0.5$ eV in (a). 
Parameters used are $U=2.232$ eV and $J=0.202$ eV for (a) and (b), and $U=1.80$ eV and $J=0.225$ eV for (c). 
Note that these bulk spectra are very similar to the ones presented in Ref. \cite{Okamoto2014} with $U=2.0$ eV and $J=0.2$ eV.  
}
\label{fig:ARPES}
\end{center}
\end{figure}

\begin{figure}
\begin{center}
\includegraphics[width=0.8\columnwidth]{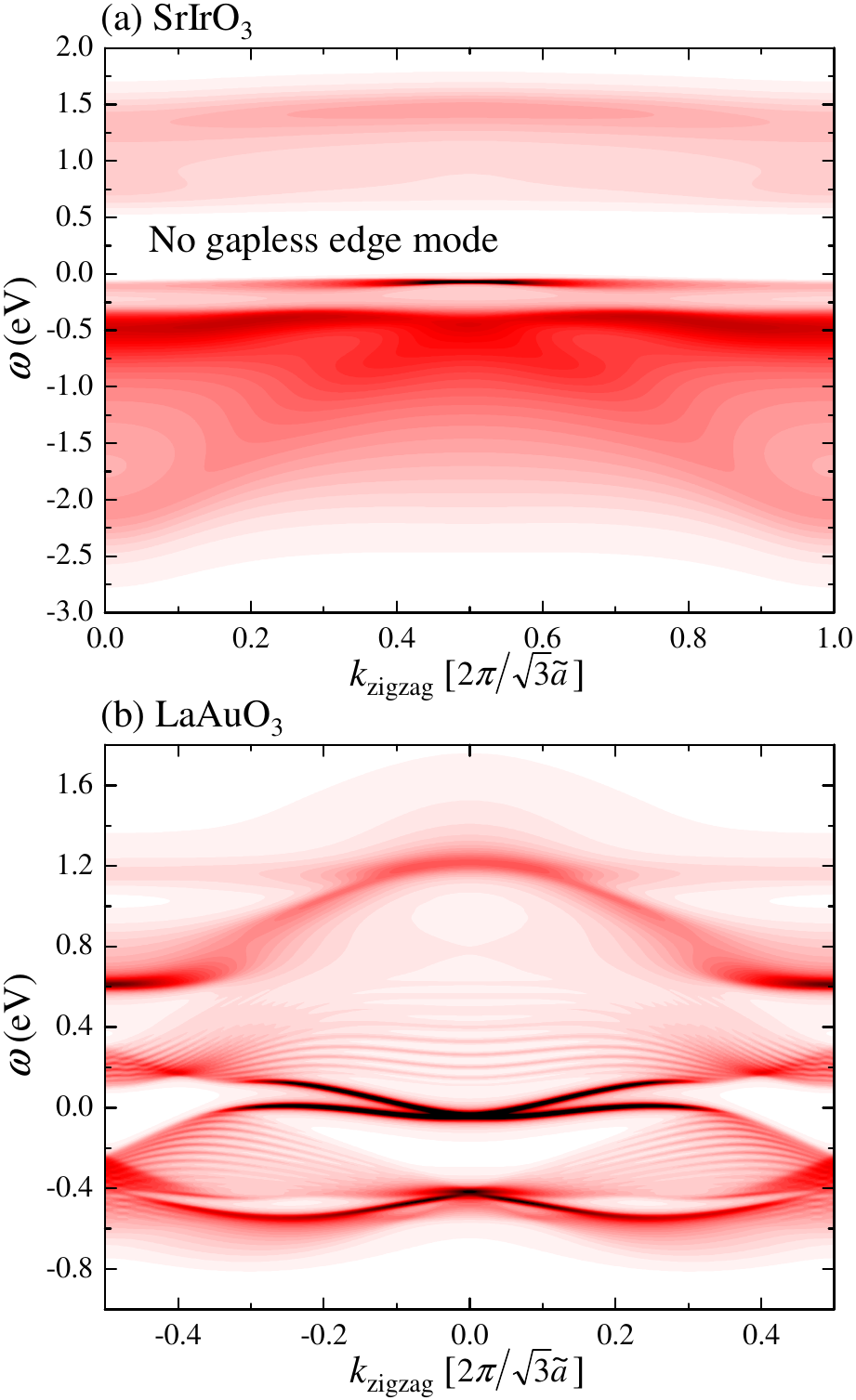}
\caption{(Color online) 
DFT+DMFT results of the ARPES spectra of the (111) bilayer of TMOs on a finite-thickness zigzag slab. 
(a) SrIrO$_3$ with the SrTiO$_3$ substrate, (b)  LaAuO$_3$ with the LaAlO$_3$ substrate.  
Parameter values are the same as in Fig. \ref{fig:ARPES}. 
Because of the AFM ordering, SrIrO$_3$ (111) bilayer is in a trivial phase and there is no gapless mode, 
while LaAuO$_3$ (111) bilayer is in  a non-trivial phase with gapless modes. 
Figures are taken from Ref. \cite{Okamoto2014}. }
\label{fig:ARPES_slab}
\end{center}
\end{figure}

Using these parameters, we found that SrIrO$_3$ (111) bilayer is unstable against AFM ordering. 
When the magnetic ordering is suppressed, it recovers a TI state, but the gap amplitude is undetectably small 
because the near Fermi energy band dominated by $j_{eff}=1/2$ state has extremely small quasiparticle weight $Z \sim 0.07$.  
The other two bands dominated by $j_{eff}=3/2$ states show only moderately renormalized quasiparticle weight  $Z \sim 0.88$ and $0.79$. 
This behavior can be seen clearly from angle-resolved spectral function in Fig. \ref{fig:ARPES} (a), where 
nearly flat bands (indicating small $Z$) appear at the Fermi level while far below the Fermi level spectral intensity nicely follows DFT dispersions. 
Such a behavior reminds us of an orbital-selective Mott transition \cite{Anisimov2002}. 
When the magnetic ordering is allowed, a fairly large gap opens [Fig. \ref{fig:ARPES} (b)]. 
On the other hand, the topological nature of LaAuO$_3$ (111) bilayer is found to be robust against correlation effects; 
magnetic ordering is unstable with the realistic parameters and the quasiparticle weight is rather large $Z \sim 0.81$ for two Kramers states.  
As shown in Fig. \ref{fig:ARPES} (c), the peak position of the spectral function and band dispersion differ only slightly. 

Noticing $J/U \approx 0.1$ in both SrIrO$_3$ and LaAuO$_3$, we also studied their systematic behavior as a function of $U$ with the fixed ratio $J/U=0.1$. 
It is found that SrIrO$_3$ (111) bilayer is indeed in the vicinity of an orbital-selective Mott transition 
with the critical interaction $U_c \sim 2.5$~eV (when $J/U=0.1$). 
Since this transition accompanies the change in the band topology from nonrivial to trivial, this might be called an orbital-selective topological Mott transition. 
This orbital-selective Mott transition does not only come from the different band width of three states dominated by $J_{eff}=1/2$ states or $J_{eff}=3/2$ states
but also from different fillings; one band dominated by $J_{eff}=1/2$ states are nearly half filled while the other bands dominated by $J_{eff}=3/2$ states 
are nearly fully filled. 
The critical interaction for the onset of AFM ordering is found to be rather small, $U_{AFM,c} \sim 0.5$~eV (when $J/U=0.1$), 
reflecting the small effective band width of $j_{eff}=1/2$-dominated band, $W \approx 0.5$ eV. 
On the other hand, LaAuO$_3$ shows a standard Mott transition where two bands undergo a mass divergence simultaneously at $U_c \sim 4.3$ eV. 
AFM ordering is also stabilized but it requires larger interaction ($U_{AFM,c} \sim 2.1$ eV) than SrIrO$_3$. 
It is worth noticing that, for LaAuO$_3$, there is a small window $2.1$ eV $\lesssim U \lesssim 2.15$ eV 
for an AFM TI where the gapless edge states and bulk magnetic ordering coexist. 
This comes from the fact that the spin component perpendicular to the [111] plane is conserved in our $e_g$ electron model, 
and therefore the LaAuO$_3$ bilayer consists of two copies of Chern insulators with up and down electrons having the opposite Chern number, 
leading to zero Hall response but quantized spin Hall response.

The trivial nature of SrIrO$_3$ and non-trivial nature of LaAuO$_3$ can be seen in the edge spectrum of finite-thickness slab. 
As shown in Fig. \ref{fig:ARPES_slab} (a), SrIrO$_3$ does not have gapless edge modes. 
On the other hand, LaAuO$_3$ has gapless edge modes [Fig. \ref{fig:ARPES_slab} (b)].

\subsection{Strong coupling approach}

In the previous subsection, we studied SrIrO$_3$ (111) bilayer and LaAuO$_3$ (111) bilayer using DFT+DMFT technique. 
We found that the SrIrO$_3$ (111) bilayer is in the vicinity of an orbital-selective Mott transition and unstable against AFM ordering. 
Once such magnetic ordering or correlation is established, a weak-coupling magnetic insulating regime and a strong-coupling magnetic insulating regime 
are adiabatically connected, thus the two approaches, weak coupling and strong coupling, would be equally valid. 
Moreover, a strong-coupling approach from a Mott insulating side could provide physically more transparent picture. 
Here, we focus on the SrIrO$_3$ (111) bilayer using strong coupling approach and discuss the origin of magnetic ordering and potential novel phenomena.\cite{Okamoto2013a} 

We start from constructing the low-energy effective Hamiltonian from 
a multiband Hubbard model with $t_{2g}$ orbitals. 
According to Wannier parameterization, typical band parameters are 
$t_\pi \sim 0.31$ eV and $t_\delta \sim 0.02$ eV for the NN hoppings, and $\Delta \sim 0.006$ eV for the trigonal crystalline field. 
Thus, for simplicity, we consider only $t_\pi$ in the following discussion.  
The effective Hamiltonian is derived from the second-order perturbation processes with respect to 
the transfer terms and by projecting the superexchange-type interactions onto the isospin states for $J_{eff}^z = \pm 1/2$: 
\begin{eqnarray}
| J_{eff}^z = \sigma \rangle = \frac{1}{\sqrt{3}} \{ i | yz, -\sigma \rangle - \sigma | xz, -\sigma \rangle + i \sigma |xy, \sigma \rangle\}. 
\end{eqnarray}
The resulting effective Hamiltonian reads 
\begin{eqnarray}
H_{KH} = \sum_{\langle \vec r \vec r' \rangle} 
\Bigl(J_K S_{\vec r}^\gamma S_{\vec r'}^\gamma +  J_H \vec S_{\vec r} \cdot \vec S_{\vec r'}  \Bigr), 
\label{eq:Hkh}
\end{eqnarray}
where, 
$\vec S$ is an isospin operator for $J_{eff} = 1/2$, not a real spin. 
$J_K$ and $J_H$ are given by 
$J_K=\frac{4}{9} (r_1 -  r_2)$ and $J_H=\frac{8}{27} (3r_1 + r_2 + 2 r_3)$, respectively, 
where 
$r_1 = t_\pi^2/(U-3J)$, $r_2 = t_\pi^2/(U-J)$, $r_3 = t_\pi^2/(U+2J)$  
with $U$ being the intraorbital Coulomb interaction and $J$ the interorbital exchange interaction [see Eq. (\ref{eq:Coulomb})]. 
$\gamma$ depends on the direction of $\vec r$-$\vec r'$ pair; i.e., 
$\gamma = x$ $(y,z)$ when the $\vec r$-$\vec r'$ bond is along the cubic $x$ $(y,z)$ axis. 
Thus, this effective Hamiltonian has the same form as the celebrated Kitaev-Heisenberg model, which is originally proposed for Na$_2$IrO$_3$.  
Here, however, both Kitaev and Heisenberg terms have a positive sign, i.e., AFM \cite{Jackeli09}. 

First we solved this model on a finite-size cluster using a Lanczos exact-diagonalization method to obtain a ground-state phase diagram. 
Following Ref. \cite{Chaloupka2010}, we parametrize $J_K$ and $J_H$ as  $J_K=2 \alpha$ and $J_H = 1-\alpha$ and vary $\alpha$ from 0 (Heisenberg limit) to 1 (Kitaev limit). 
In contrast to a FM-Kitaev/AFM-Heisenberg model, a model proposed to describe Na$_2$IrO$_3$ in Ref. \cite{Chaloupka2010}, 
there appear two phases in AFM-Kitaev/AFM-Heisenbeg model $H_{KH}$: 
an AFM N{\' e}el ordered phase at $\alpha < \alpha_c$ and a Kitaev spin-liquid phase at $\alpha > \alpha_c$ with $\alpha_c \sim 0.96$. 

For a general Kitaev-Heisenberg model, where $J_K$ and $J_H$ could be either FM or AFM, 
a more complex phase diagram is obtained, including a zigzag AFM phase and a FM phase \cite{Okamoto2013b,Chaloupka2013}. 

Now, a question arises, {\it where would be the realistic range for SrIrO$_3$ (111) bilayer in this phase diagram?} 
Note that $U>3J$ is physical because, otherwise, the charge excitation energy from $t_{2g}^5 t_{2g}^5$ to $t_{2g}^6 t_{2g}^4$ becomes negative. 
From the analytic expressions of $J_K$ and $J_H$, the $J_K/J_H$ ratio is an increasing function of $J$; 
$\alpha $ varies from 0 to 1/5 when $J$ is varied from 0 to $U/3$. 
So, the physical parameter range would be $\alpha < 1/5$, where a N{\'e}el AFM ordering is stabilized. 
This result is consistent with our DFT+DMFT predictions. 

While the novel Kitaev spin liquid phase exists in our effective model, 
realizing it using SrIrO$_3$ (111) bilayer seems to be very difficult, even though it may not be entirely impossible. 
However, carrier doping may be feasible. 
This might induce novel phenomena as carrier doping into a Mott insulator on a square lattice induces $d$-wave superconductivity \cite{Baskaran1987,Lee2006}. 

In addition to the effective spin model Eq. (\ref{eq:Hkh}), we consider hopping matrices projected into neighboring or $J_{eff}^z = \pm 1/2$ states. 
In this representation, the hopping matrices are diagonal in the isospin index $\sigma$ and given by 
$H_t = - \tilde t \sum_{\langle \vec r \vec r' \rangle \sigma} (\tilde d_{\vec r \sigma}^\dag \tilde d_{\vec r' \sigma} + H.c.)$ 
with the renormalized hopping amplitude $\tilde t = \frac{2}{3} t_\pi$. 
$\tilde d_{\vec r \sigma}$ excludes the double occupancy because of the strong Coulomb repulsive interaction. 
Thus, the effective model for the doped Kitaev-Heisenberg model is given by $H_{eff}=H_t + H_{KH}$. 
We analyze this model using a $SU(2)$ slave boson mean field (SBMF) method developed for high-$T_c$ cuprates \cite{Lee2006}. 
Because of the Kitaev interaction, there appears a larger number of mean field order parameters than for the $t$-$J$ model. 
To make the problem tractable, we focus here on several ans{\"a}tze which respect the sixfold rotational symmetry of the underlying lattice. 
The detail of these ans{\"a}tze and the mean field calculations can be found in Ref. \cite{Okamoto2013a}. 

\begin{figure}
\begin{center}
\includegraphics[width=0.9\columnwidth]{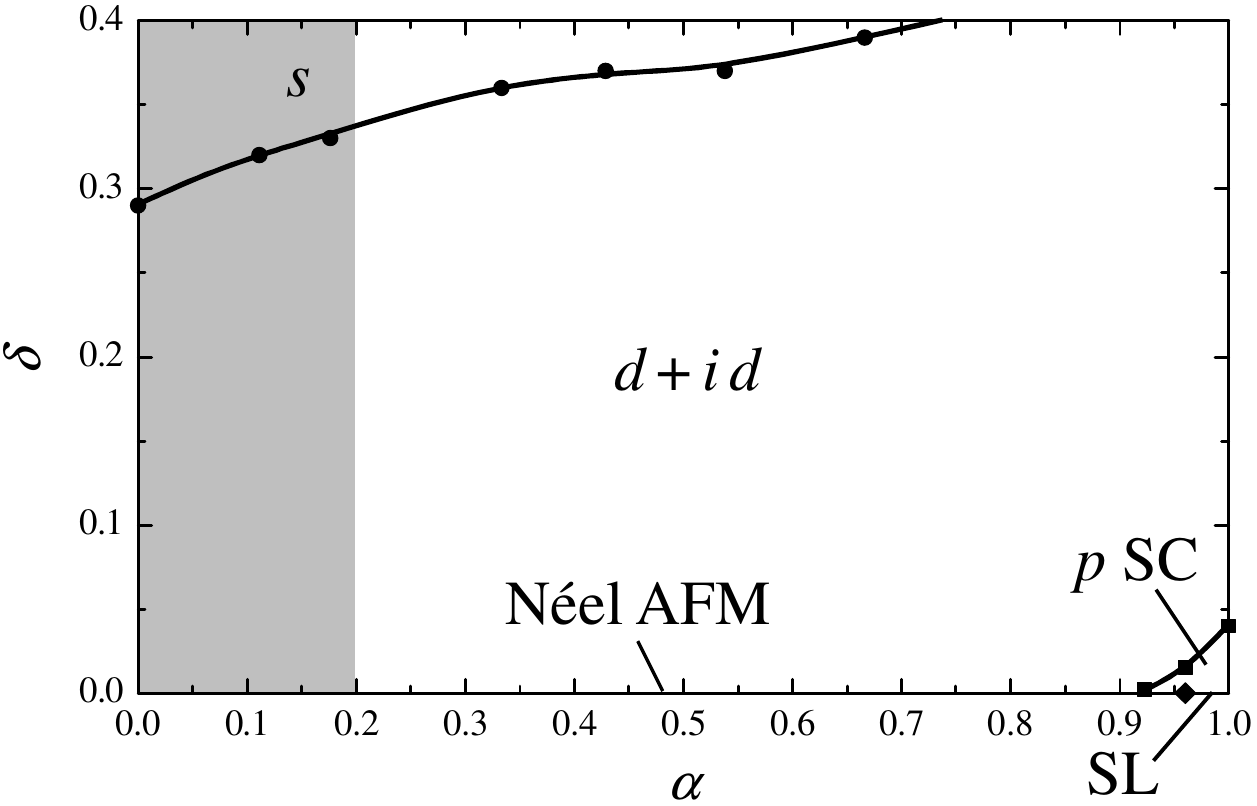}
\caption{
Schematic phase diagrams for the doped AFM-Kitaev/AFM-Heisenberg model as a function of $\delta$ and $\alpha$. 
Parameters are taken as $J_K+J_H=t_\pi $. 
$\alpha$ is the measure of the relative strength between $J_K$ and $J_H$ as $J_K=2 \alpha$ and $J_H=1-\alpha$.
Phase boundaries at finite $\delta$ are the results of the $SU(2)$ SBMF, 
while those at $\delta=0$ are results of the exact diagonalization. 
The shaded area is the realistic parameter range for $0<3J<U$. 
The figure is taken from Ref. \cite{Okamoto2013a}.}
\label{fig:kitaev}
\end{center}
\end{figure}

The mean field phase diagram as a function of $\alpha$ and doping concentration $\delta$ is shown in Fig. \ref{fig:kitaev}. 
The shaded area is the realistic parameter range. 
Since the Heisenberg interaction dominates in the undoped limit, the effect of carrier doping is also similar to 
the $t$-$J$ model on a honeycomb lattice, and  
a large area is covered by a superconducting (SC) state with the $d+id$ ($d_{x^2-y^2}+id_{xy}$) pairing \cite{BlackSchaffer2007}. 
Similar to this result, carrier doping to a square-lattice iridium oxide Sr$_2$IrO$_4$ is 
theoretically suggested to induce the $d_{x^2-y^2}$ SC.\cite{Wang2011a,Watanabe2013,Meng2014} 
In addition, we found a $s$-wave SC state at a large $\delta$ regime and a $p$-wave SC state in the vicinity of the Kitaev spin liquid phase. 
The relative stability between the $d+id$ and $s$ wave SC states is determined by the competition between the pairing formation and the coherent motion of carriers. 
The $p$-wave SC state is unique for Kitaev interactions.
This breaks the TRS and is adiabatically connected to the undoped Kitaev spin liquid state.
A similar $p$-wave SC state is also found in a doped FM Kitaev model in Ref. \cite{You2011}, 
while Ref. \cite{Hyart2012} found another $p$-wave SC state which does not break the TRS. 
These two $p$-wave SC states are found to be separated by a first order transition by changing $\delta$ \cite{Okamoto2013b}. 
Ref. \cite{Okamoto2013b} also studied a general Kitaev-Heisenberge model with finite $\delta$  and reported a complex phase diagram. 

While the (111) bilayer of SrIrO$_3$ may not be sufficient for studying the novel physics associated with Kitaev interactions, 
carrier doping or replacing Ir with some other elements may allow the access to rich behavior as discussed in this subsection.  
 
\section{Related work}

After our first proposal \cite{Xiao2011}, a number of related works have appeared.\cite{Liu2016} 
Here, we list some of the important ones. 

{\it SrIrO$_3$}: 
Theoretical work based on the DFT calculation, similar to ours, has appeared in Ref. \cite{Lado2013}. 
The experimental realization was quite challenging because of the large lattice mismatch between SrIrO$_3$ and SrTiO$_3$. 
In Ref. \cite{Hirai2015}, Hirai and coworkers were successful to fabricate a bilayer of iridate by replacing a half of Sr by Ca, 
i.e., Ca$_{0.5}$Sr$_{0.5}$IrO$_3$, on SrTiO$_3$. 
They found such an iridate (111) bilayer is strongly insulating and magnetic. 
While the detail of magnetic ordering is not clarified yet, the trivial insulator nature with magnetism is consistent with our DFT+DMFT results. 

{\it LaAuO$_3$ and LaAgO$_3$}: 
Similar to our proposal, LaAuO$_3$ and LaAgO$_3$ (111) bilayers were also theoretically studied. 
In Ref. \cite{Liang2013}, Liang and Hu proposed growing those TMOs between LaCrO$_3$ to realize controllable Chern insulators. 

{\it LaNiO$_3$}: LaNiO$_3$-based superlattices have been actively studied because of its potential to realize the unconventional superconductivity similar to cuprates.\cite{Chaloupka2008} 
(111) bilayers of nickelates were first studied using model Hamiltonians to predict novel anomalous Hall insulators due to the complex orbital ordering.\cite{Ruegg2011,Yang2011}
This orbital ordering accompanies a finite expectation value of $\langle \tau^y \rangle$. 
Thus, the SOC-like effect is dynamically generated. 
Later, realistic DFT calculations were performed.\cite{Ruegg2012,Ruegg2013,Doenning2014}
Experimental efforts on such nickelate (111) bilayer have also appeared.\cite{Middey2012,Middey2016} 

{\it La(Ni,Co)O$_3$}: Correlation induced novel phenomena have been theoretically predicted, including 
odd-parity superconductivity in a (111) bilayer of LaNi$_{7/8}$Co$_{1/8}$O$_3$ \cite{Ye2014} 
and quantum anomalous Hall phases in a (111) bilayer of LaCoO$_3$.\cite{Wang2015}

{\it SrTiO$_3$}: SrTiO$_3$/LaAlO$_3$ heterostructures are one of the most intensively studied systems due to the high-mobility electron gases.\cite{Ohtomo2004} 
In Ref. \cite{Doenning2013}, (111) bilayers of SrTiO$_3$ grown on LaAlO$_3$ were theoretically studied. 

{\it LaMnO$_3$}: 
When $e_g$ band is fully spin polarized, the Fermi level for an $e_g^1$ system could be just inside the nontrivial gap [see Fig. \ref{fig:TB} (c)]. 
Since the spin quantum number along the [111] direction is conserved, a resulting electronic phase could be a Chern insulator. 
Such a state was indeed theoretically reported in LaMnO$_3$ (111) bilayers.\cite{Weng2015,Doennig2016}

{\it Other $4d$ and $5d$ perovskite TMOs}: 
There have also appeared a number of theoretical studies on $4d$ and $5d$ perovskite TMOs. 
Among these TMOs, Refs.~\cite{Chandra2017,Guo2017}showed that LaOsO$_3$ (111) bilayer could become a Chern insulator with high Curie temperature. 
Ref.~\cite{Guo2017} also showed a similar Chern insulating state in LaRuO$_3$ (111) bilayer. 
Further, Ref.~\cite{Si2017} showed that SrRuO$_3$ (111) bilayer could also show a Chern insulating state when electrons are doped. 
%
Ref.~\cite{Yamauchi2015} considered BiIrO$_3$ (111) bilayers sandwiched by BiAlO$_3$. 
Since an Ir ion has the $t_{2g}^6$ configuration, this system has a trivial band topology. 
Nevertheless, novel spin-valley coupled phenomena\cite{Xiao2012} could result. 

{\it Double perovskite}: 
When the top $B$ sites and the bottom $B$ sites of $AB$O$_3$ (111) bilayer are different, it is a (111) monolayer of double perovskite. 
In Ref. \cite{Cook2014}, such heterostructures were studied using Fe on the top layer and Os on the bottom layer 
to predict novel Chern insulating states. 

{\it Dice lattice}: When a trilayer of $AB$O$_3$ is grown along the (111) direction, $B$ sites form a ``dice'' lattice. 
Ref. \cite{Wang2011} discussed possible Chern insulators associated with nearly flat bands induced in such a dice lattice. 

{\it Pyrochlore oxides}: In addition to perovskite oxides, pyrochlore oxides have been studied theoretically. 
Since the (111) plane is the natural cleavage plane, this class of materials might be grown along the [111] direction more easily. 
Novel topological phases have been predicted using model Hamiltonians \cite{Hu2012,Treacher2012,Yang2014,Bergholts2015}
and the first principle DFT calculations.\cite{Hu2015}
Further, topological magnon states and unconventional superconductivity in pyrochloe iridates are discussed in Ref. \cite{Laurell2017}. 

{\it Corundum and other hexagonal TMOs}: 
As mentioned in Ref. \cite{Xiao2011}, corundum materials are possible candidates for realizing novel quantum states associated with honeycomb lattice 
because the [0001] plane of corundum $M_2$O$_3$ involves a honeycomb lattice formed by $M$ atoms.
Ref. \cite{Afonso2015} reported nontrivial band topology using DFT techniques with $M= {\rm Au}$ and Os grown between sapphire Al$_2$O$_3$. 

Similar to the aforementioned Na$_2$IrO$_3$, SrRu$_2$O$_6$ has a honeycomb lattice formed by Ru ions. 
However, because of the robust G-type AFM ordering with relatively high N{\'e}el temperature $T_N \sim 565$ K,\cite{Hiley2014,Tian2015,Hiley2015} 
its topological property has not attracted attentions. 
A recent theoretical study showed that  the band topology of SrRu$_2$O$_6$ is indeed trivial even when the AFM ordering is suppressed 
but this could be turned to nontrivial by a slight modification of lattice parameters \cite{Ochi2016}. 
While realizing nontrivial band topology in SrRu$_2$O$_6$ remains challenging, 
it was suggested that the topological transition could be induced under accessible pressures by replacing Sr by Ca and Ru by Os,\cite{Ochi2016} 
resulting in a strong TI.\cite{Fu2007} 
Although this is not a heterostructure, 
thin film-growth techniques developed for the TMO heterostructures could make important contributions 
for growing such materials and introducing pressure or strain induced by a substrate.

\section{Conclusion}

In this article, we reviewed our theoretical work on (111) bilayers of perovskite TMOs as a fertile playground to explore novel electronic phenomena. 
Our work started from an extremely simple idea; honeycomb-like structure could be realized using perovskite TMOs when grown along the [111] direction, 
which is almost like {\it an egg of Columbus}.  
We found surprisingly rich behavior from (111) bilayers of perovskite TMOs, including 
2-dimensional TIs even for $e_g$ electron systems in which the SOC is quenched in bulk systems, orbital-selective Mott transition, 
Kitaev physics, and doping-driven unconventional superconductivity. 
Beside our original proposal, there has appeared much research on (111) TMO heterostructures. 
While an experimental realization of truly novel or topological phenomena as theoretically predicted has yet to appear, 
we hope that further developments in the thin-film growth technique will eventually realize some of the theoretical predictions 
and our work will make important contributions to modern materials science. 







\begin{acknowledgments}


The research by S.O. is supported by 
the U.S. Department of Energy,  Office of Science, Basic Energy Sciences, Materials Sciences and Engineering Division.
D.X. acknowledges support by the Air Force Office of Scientific Research under Grants No. FA9550-12-1-0479. 
We would like to thank 
W. Zhu, Y. Ran, N. Nagaosa, Y. Nomura, and R. Arita for collaborations and fruitful discussions at every stage of this work. 

\end{acknowledgments}






\end{document}